# $CO_2$ dissociation activated through electron attachment on reduced rutile $TiO_2(110)$-1×1 surface

Shijing Tan[a)], Yan Zhao[a)], Jin Zhao, Zhuo Wang, Chuanxu Ma, Aidi Zhao, Bing Wang*, Yi Luo, Jinlong Yang, Jianguo Hou*

*Hefei National Laboratory for Physical Sciences at Microscale, University of Science and Technology of China, Hefei, Anhui 230026, P.R. China*

Converting $CO_2$ to useful compounds through the solar photocatalytic reduction has been one of the most promising strategies for artificial carbon recycling. The highly relevant photocatalytic substrate for $CO_2$ conversion has been the popular $TiO_2$ surfaces. However, the lack of accurate fundamental parameters that determine the $CO_2$ reduction on $TiO_2$ has limited our ability to control these complicated photocatalysis processes. We have systematically studied the reduction of $CO_2$ at specific sites of the rutile $TiO_2(110)$-1×1 surface using scanning tunneling microscopy at 80 K. The dissociation of $CO_2$ molecules is found to be activated by one electron attachment process and its energy threshold, corresponding to the $CO_2^{\bullet-}/CO_2$ redox potential, is unambiguously determined to be 2.3 eV higher than the onset of the $TiO_2$ conduction band. The dissociation rate as a function of electron injection energy is also provided. Such information can be used as practical guidelines for the design of effective catalysts for $CO_2$ photoreduction.

During the last decades, there is a growing research interest in converting $CO_2$ into value added products for energy production to actively reduce the $CO_2$ emission.[1-12] One of the promising strategies is to convert $CO_2$ into CO or hydrocarbons by photoreduction,[13-17] although its efficiency still needs to be significantly improved.[18] The decisive step in the $CO_2$ reduction is to effectively generate $CO_2^{\bullet-}$, the electron attached state of $CO_2$,[19,20] which is controlled by the reduction potential of the $CO_2^{\bullet-}/CO_2$ redox couple. The search for a good match between the reduction potential and the conduction band (CB) of the photocatalytic substrates has been the central focus of many studies. It is found that even for the widely used photocatalyst, $TiO_2$, a strong mismatch occurs,[3,5,20-23] resulting in highly unfavorable electron transfer from the photo-excited conduction band of the $TiO_2$ to the $CO_2$. Such energy mismatch could be compensated by either introducing addition catalysts to assist the electron transfer or modifying the conduction band of the photocatalyst with chemical modifications.[5] However, the optimization procedures are hampered by the lack of accurate data for the bonding sites of $CO_2$ on the substrates and the energy position of the reduction potential. In this case, an atomistic study with scanning tunneling microscopy (STM) is highly desirable since it can not only provide a complete picture for the specific adsorption sites of single $CO_2$ molecules, but also be able to determine the reduction potential through the detection of the unoccupied molecular orbitals. Here, we present a comprehensive study on STM induced one-step direct reduction process of $CO_2$ to CO on reduced rutile $TiO_2(110)$-1×1 surface at 80K. The adsorption sites of $CO_2$ at various coverages, the reduction potential of the $CO_2^{\bullet-}/CO_2$ redox couple and the reaction rate are accurately determined. The underlying mechanisms are fully examined with the help of first principles calculations.

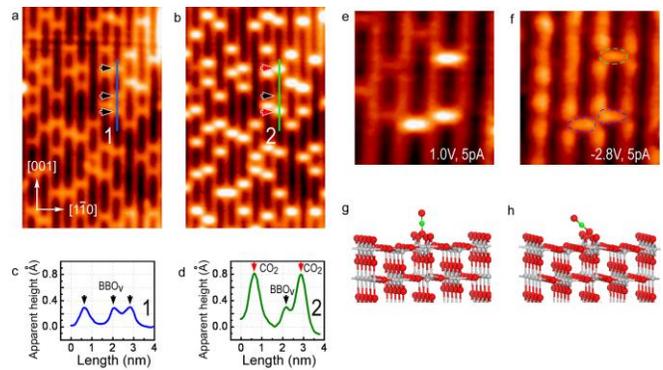

**Fig. 1 a**, **b**, Images of $TiO_2(110)$-1×1 before and after in situ $CO_2$ adsorption at 80 K. (Size: 7.2 × 11.4 nm², Imaging conditions: 1.0 V, 10 pA). **c**, **d**, Line profiles showing the apparent height of adsorbed $CO_2$ in comparison with the $BBO_V$. **e**, **f**, Unoccupied-state (1.0 V) and occupied-state (-2.8 V) images of adsorbed $CO_2$ on $TiO_2$, respectively (Size: 2.9 × 3.2 nm²). The protrusions for the adsorbed $CO_2$ in occupied-state image show symmetric and asymmetric shapes, marked by the dashed ellipse and spindly oval in **f**, respectively. **g**, **h**, Structures of adsorbed $CO_2$ with vertical and inclined configurations.

Figure 1 shows the STM images within the same area of hydroxyl-free $TiO_2(110)$-1×1 surface before and after the exposure of 0.15 langmuir $CO_2$ (1 langmuir = 1×10⁻⁶ Torr·s) at 80 K. After the $CO_2$ exposure, it is observed that the $CO_2$ molecules only appear at the bridge-bonded oxygen vacancy ($BBO_V$) sites, as the protrusions shown in Figure 1b. The apparent height of $CO_2$ is about 0.8Å (Figure 1c and 1d). Differently, CO preferentially adsorbs at $Ti^{4+}$ site close to a $BBO_V$ but not directly at the $BBO_V$.[24] $CO_2$ molecules adsorbed at BBO$_V$ both show protrusions either in the unoccupied-state or occupied-state images, but less protruded in the latter case (Figure 1e and 1f). By varying the CO$_2$ coverage, we found that the CO$_2$ could appear at the Ti$^{4+}$ sites only after all the BBO$_V$s were completely filled by CO$_2$. With the excess exposure of CO$_2$, the diffusive CO$_2$ may occur at the Ti$^{4+}$ site. (see Figure S1 in the supporting materials), but no stable adsorption configuration can be imaged, even at a much lower temperature of 15 K. (see Figure S2 in the supporting materials). Our density functional theory (DFT) calculations indicate that the CO$_2$ linearly adsorbs at the BBO$_V$ site with vertical and inclined configurations, as schematically shown in Figure 1g and 1h, respectively. The symmetric and asymmetric shape of adsorbed CO$_2$ in the occupied-state image can be attributed to the different adsorption configurations, as indicated in Figure 1f. The adsorption energies are estimated to be 0.50 and 0.76 eV, respectively with vertical and inclined configurations at BBO$_V$, much larger than the value of 0.17 eV for the CO$_2$ on Ti$^{4+}$ site with the O–C–O bond parallel to [001] direction, in agreement to the previous TPD results that the CO$_2$ binds to the BBO$_V$ of Ti$^{3+}$ sites more strongly than to fivefold coordinated Ti$^{4+}$ sites.[25-30] However, our STM observations do not support recent reported results that the stable adsorption configuration of CO$_2$ presents at the Ti$^{4+}$ site.[31,32] We believe their observed bright protrusions at the Ti$^{4+}$ site measured at 80 K could be resulted from species other than adsorbed CO$_2$ molecules.

Figure 2a-c give a set of STM images showing the STM tip induced CO$_2$ dissociation. It is found that the adsorbed CO$_2$ molecules at BBO$_V$ can be removed when a relatively high voltage pulse is applied by the tip (Figure 2b-c). By comparing Figure 2c with Figure 2a, it is clearly demonstrated that together with the disappearance of the CO$_2$ molecules, the original BBO$_V$s also disappear. This strongly indicates that the CO$_2$ is actually dissociated into an oxygen atom and a CO molecule. At the Ti$^{4+}$ site shown at the lower-right of Figure 2c, the observed protrusions after the CO$_2$ dissociation are quite different from that with the adsorbed CO$_2$, but fit well with the CO adsorption behavior as we observed before.[24] Evidently, they are the readsorbed CO molecules from the dissociation product of the CO$_2$. Therefore, in the dissociation process, the oxygen atom occupies the BBO$_V$ vacancy, while the CO molecule either desorbs from the surface or adsorbs at Ti$^{4+}$ site, as schematically shown in Figure 2e. A typical current-time (*I-t*) curve is given in Figure 2d, recorded during applying the voltage pulse. The current jump in the *I-t* curve reflects the dissociation of CO$_2$, which could be used to measure the dissociation rate of the CO$_2$. The plot of the tip-induced dissociation rate as a function of tunneling current at different bias voltages is given in Figure 2f. The dependence on current is linear and yields a slope in the log-log plot of 0.98 ± 0.10 (2.6 V), 0.96 ± 0.07 (2.4 V), and 1.05 ± 0.01 (2.2 V). These values clearly imply that the

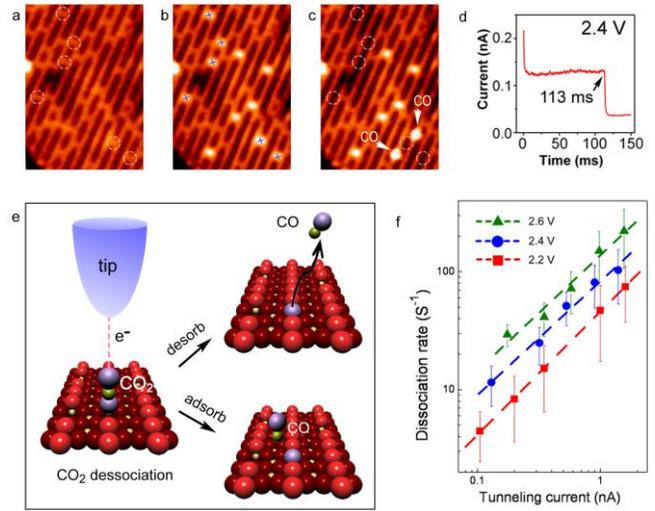

**Fig. 2 a**, Images of bare TiO$_2$(110)-1×1 surface, **b**, after CO$_2$ adsorption in situ at 80 K, **c**, after tip-induced CO$_2$ dissociation. (Size: 8.5 × 11.9 nm$^2$, Imaging conditions: 1.0 V, 10 pA). **d**, A typical *I-t* curve during the voltage pulse. **e**, Schematic drawing of the tip-induced CO$_2$ dissociation, leading to the healing of the BBO$_v$ and either desorbed CO or adsorbed CO at Ti$^{4+}$ site. **f**, Plot of CO$_2$ dissociation as a function of the tunneling current measured at different bias voltages.

dissociation process involves only one electron per dissociation event, ruling out a nonlinear ''vibrational heating'' mechanism.[33,34] This situation happens only if the tunneling electrons are directly injected into the lowest unoccupied molecular orbital (LUMO) of the adsorbed CO$_2$. It can thus be used to determine the surface state of the CO$_2$ upon adsorption.

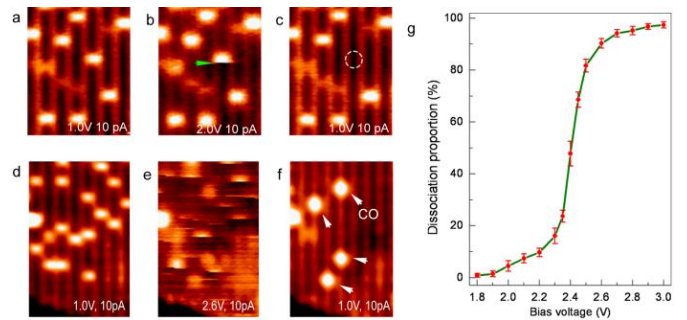

**Fig. 3 a-c,** a set of images showing CO$_2$ dissociation during scanning at 2.0 V (Size: 3.9 × 4.7 nm$^2$). **d-f**, another set of images showing CO$_2$ dissociation during scanning at 2.6 V (Size: 5.1 × 8.5 nm$^2$). **g**, Dissociation proportion of adsorbed CO$_2$ at BBO$_V$ as a function of the applied bias voltage.

Figure 3a-c and 3d-f show two sets of images during the CO$_2$ dissociation period at relatively high bias voltages

of 2.0 and 2.6 V. It is found that when the surface is scanned with a bias voltage of 2.0 V, only a few of $CO_2$ can be dissociated, as marked by the arrow in Figure 3b. With a higher bias voltage of 2.6 V, nearly all of the $CO_2$ could be dissociated within the scanning area, accompanying with disappearance of the $BBO_{VS}$ and with occurrence of some CO at the $Ti^{4+}$ sites, as shown in Figure 3d-f. We have not found the diffusion of the adsorbed $CO_2$ at $BBO_V$ even under relatively high bias voltages. The dissociation population as a function of the bias voltage is plotted in Figure 3g. A distinct increase of the population appears when the applied bias voltage is higher than 2.3 V, and almost all of the adsorbed $O_2$ molecules can be dissociated during only one scan when the voltage is higher than 2.6 V. It is noted that no dissociation event can be detected when the bias voltage is below 1.8 V, which could be regarded as the voltage threshold for the dissociation of the $CO_2$. The threshold is not dependent on the set point current in the range of 10 pA and 10 nA. We also tried to dissociate the $CO_2$ using the negative bias voltages from −1.8 to −4.0 V, but did not observe any dissociation events. This means the dissociation of $CO_2$ can only happen by the electron injection. We also performed the experiment by illuminating the $CO_2$ adsorbed sample with UV light and pulsed laser of wavelength of 266 nm, with the method described elsewhere,[35] but failed to observe any $CO_2$ dissociation events, although some of the $CO_2$ may hop between $BBO_{VS}$ (see Figure S3 in the supporting materials).

The tip-induced dissociation of molecules on solid surfaces using STM has been observed in other molecular systems.[33,34,36-41] Generally, such dissociation is attributed to the inelastic tunneling electrons (IETE) that induce vibrational excitations[33] or electronic excitation[42] of the adsorbed molecules. In this case, IETE services as an energy source and the total number of electrons in the molecule remains the same during the dissociation process. Our experimental results have strongly indicated that the $CO_2$ dissertation is most likely to be a one-step reduction process, in which a tunneling electron is attached to the $CO_2$ molecule to form $CO_2^{\bullet-}$. To confirm the hypothesis and understand our experimental results, we have carried out first principles calculations to examine the interaction of the adsorbed $CO_2$ with $TiO_2$ surface and to determine the actual surface states of $CO_2$ upon adsorption on $TiO_2$ that control the electron attachment process.

Figure 4a gives the calculated partial density of states (PDOS) of the adsorbed $CO_2$ at $BBO_V$. It is observed that the energy gap of the adsorbed $CO_2$ almost maintains that of the free $CO_2$ in both of the vertical and inclined adsorption configurations. The LUMO locates above the conduction band (CB) onset of $TiO_2$ by 2.3 eV for both vertical and inclined adsorption configurations and it mixes with the 3d orbital of $Ti^{3+}$ of the $BBO_V$ (Figure 4b). From the energetic point of view, under the photo-excitation, it is difficult for the excited electron to transfer from the CB of $TiO_2$ to $CO_2$ because of fast relaxation. This explains why

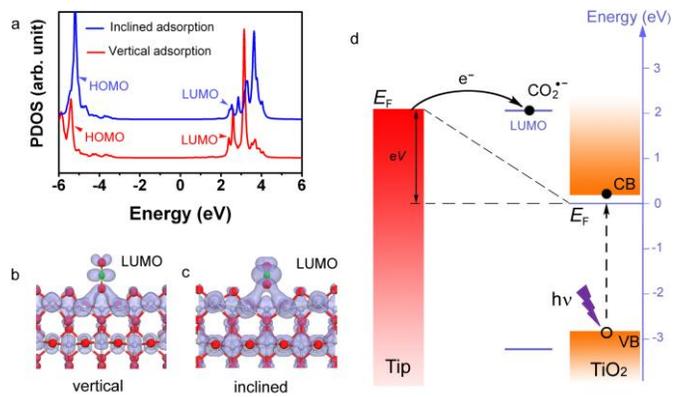

**Fig. 4 a**, PDOS of adsorbed $CO_2$ at $BBO_V$ with vertical and inclined configurations, respectively. The PDOS is shifted for clarity. **b**, Charge density distribution of LUMO in both of the vertical and inclined adsorption configurations. **c**, Illustration of the formation of $CO_2^{\bullet-}$ excited state through tunneling electron attachment.

the one step reduction of $CO_2$ could not be observed under the photoexcitation. The location of the LUMO fits well with the experimental fact that the rapid increase of the dissociation proportion around 2.3 V (Figure 3g). At this voltage, the direct injection of the electron into the LUMO of $CO_2$ molecule is thus highly feasible. As illustrated in Figure 4c, the tunneling electron can be attached to the $CO_2$ through the LUMO, forming activation state of $CO_2^{\bullet-}$. Adopting the measured threshold voltage for $CO_2$ dissociation, it is reasonable to conclude that the $CO_2^{\bullet-}/CO_2$ surface state at the solid-gas interface of $CO_2$-$TiO_2$ locates at 2.3 eV above the Fermi level. Since in our experimental setup the $TiO_2$ is of n-type and its Fermi level is close to the CB,[43] one can thus roughly estimate that the $CO_2^{\bullet-}/CO_2$ surface state should be 2.3 eV higher than the CB onset. This value is much smaller than the estimated value of 3.5 eV by Indrakanti et al,[3] but larger than the value of 1.6 eV (or −1.9 V vs SHE) in aqueous solution.[23] The accurate determination of surface state will certainly be useful for correct design effective catalysts.

In summary, we have studied the adsorption behavior of $CO_2$ molecules on $TiO_2$(110)-(1×1) surfaces using in-situ STM at 80 K. Our findings suggest that the $CO_2$ adsorbs on the top of $BBO_V$ at low coverage and the $CO_2$ dissociation is induced by the attachment of the tunneling electron from the tip. Such a hypothesis is confirmed by first principles calculations. With STM experiments, the exact location of the surface state that contributes to the formation of the $CO_2$ radical can be firmly determined, which helps to understand the preconditions for the photo-excitation process and to find ways to the improve the efficiency for the conversion of $CO_2$ into CO and other carbonyl compounds, such as methanol synthesis and methane production.

**Methods.** The STM experiments were conducted with a low temperature scanning tunneling microscope (Matrix, Omicron) in an ultra-high vacuum system with a base pressure less than $3\times10^{-11}$ mbar, which has been baked out sufficiently for a long time to minimize the background water in the chamber. All of the STM measurements were performed at 80K. An electrochemically etched polycrystalline tungsten tip was used in all STM experiments. The rutile $TiO_2$ (110) sample (Princeton Scientific Corporation) was prepared by repeated cycles of ion sputtering (3000 eV $Ar^+$) and annealing (at 900 K). The $CO_2$ gas (purity of 99.999%, Nanjing Shangyuan Industrial Gas) was used.

**Theoretical calculations.** A $TiO_2$ (110)-1×1 surface was modeled by periodically repeated slabs consisting of a (6×2) cell with 5 O-Ti-O layers separated by 10Å of vacuum. All the calculations are performed with the Vienna ab initio simulation package (VASP) with the generalized gradient approximation of Perdew, Burke, and Ernzerhof (PBE-GGA).[44-47] A plane-wave basis set with energy cutoff of 400 eV and the projector augmented wave (PAW) potential was employed.[48] Monhkorst-3 Pack grids of (2×2×1) K-points were used for the (6×2) unit cells. During the optimization, atoms were allowed to relax in the upper 3 layers and all the structures are relaxed until self-consistent forces are smaller than 0.02 eV/Å.

**Acknowledgements** This work was supported by NBRP (grant 2011CB921400, 2010CB923300) and NSFC (grants 9021013, 10825415, 10874164, 21003113), China.